\documentclass[conference]{IEEEtran}
\IEEEoverridecommandlockouts
\usepackage{cite}
\usepackage{amsmath,amssymb,amsfonts}
\usepackage{algorithmic}
\usepackage{graphicx}
\usepackage{textcomp}
\usepackage{xcolor}

\usepackage{multirow}
\usepackage{booktabs}
\usepackage{makecell}

\def\BibTeX{{\rm B\kern-.05em{\sc i\kern-.025em b}\kern-.08em
    T\kern-.1667em\lower.7ex\hbox{E}\kern-.125emX}}
\begin{document}

\title{Graph-VQE: A CUDA-Q Multi-QPU Simulation Framework for Hamiltonian-Aware Protein-Folding VQE}

\author{
\centering
\IEEEauthorblockN{Yujun Feng}
\IEEEauthorblockA{\textit{Miami University}\\
Oxford, OH, United States \\
fengy46@miamioh.edu}
\and
\IEEEauthorblockN{Yuqi Zhang}
\IEEEauthorblockA{\textit{Kent State University}\\
Kent, OH, United States \\
yzhan135@kent.edu}
\and
\IEEEauthorblockN{Jingyi Huang}
\IEEEauthorblockA{\textit{Miami University}\\
Oxford, OH, United States \\
huangj84@miamioh.edu}
\and
\IEEEauthorblockN{Bo Fang}
\IEEEauthorblockA{\textit{University of Texas at Arlington}\\
Arlington, TX, United States \\
Bo.fang@uta.edu}
\and
\IEEEauthorblockN{Shuai Xu}
\IEEEauthorblockA{\textit{Case Western Reserve University}\\
Cleveland, OH, United States \\
sxx214@case.edu}
\and
\IEEEauthorblockN{Qiang Guan}
\IEEEauthorblockA{\textit{Kent State University}\\
Kent, OH, United States \\
qguan@kent.edu}
\and
\IEEEauthorblockN{Yang Zhang}
\IEEEauthorblockA{\textit{Miami University}\\
Oxford, OH, United States \\
zhang981@miamioh.edu}
}

\maketitle

\begin{abstract}
The Variational Quantum Eigensolver (VQE) is essential for molecular simulation in drug discovery, but hardware noise and algorithmic limits restrict its precision. While the NVIDIA CUDA-Q platform mitigates some hardware issues via exact simulation, it lacks Qiskit support and restricts parallelization. To solve this, we introduce Graph-VQE, a novel framework that extends CUDA-Q with optimization-level parallelism. Graph-VQE leverages amino acid sequence structures by partitioning Hamiltonian interaction graphs into weakly coupled clusters using Louvain community detection. These clusters undergo restricted updates on the full-Hamiltonian objective, followed by a global refinement stage utilizing Hamiltonian batching. Furthermore, a custom Qiskit-CUDA-Q integration layer enables standard workflows with GPU acceleration. Evaluations on protein folding tasks prove that Graph-VQE outperforms baselines, achieving lower final energies. It delivers competitive RMSD and binding affinity compared to AlphaFold3 and IBM quantum processors while maintaining stable quality across multi-GPU environments, thereby providing a highly practical path toward high-fidelity biomolecular simulations.
\end{abstract}


\section{Introduction}

The Variational Quantum Eigensolver (VQE) is a leading hybrid quantum-classical algorithm for molecular simulation, enabling ground-state energy estimation critical for drug discovery, protein structure prediction, and materials design \cite{peruzzo2014variational, mcclean2016theory}. However, current quantum hardware suffers from gate errors, decoherence, and limited circuit depth, preventing VQE from reaching the ``chemical accuracy'' (1 kcal/mol) required for reliable predictions \cite{preskill2018quantum, o2016scalable, helgaker2013molecular}. Beyond hardware noise, standard VQE also faces an algorithmic bottleneck: the classical optimizer updates a single global parameter vector per iteration, and this sequential process becomes increasingly expensive as molecular systems grow, both because the number of tunable parameters grows and the optimization landscape flattens at scale, making it increasingly difficult for the optimizer to find directions of improvement \cite{tilly2022variational, mcclean2018barren}. Parallelizing the optimization itself, rather than only the measurement step, is therefore essential to scale VQE toward biologically relevant systems such as protein folding, where identifying the native state demands absolute energy minimization and failure produces misfoldings linked to diseases like Alzheimer's and Parkinson's \cite{dill2012protein, emani2021quantum}.

\begin{figure}[htbp]
  \centering
  \resizebox{\linewidth}{!}{\includegraphics{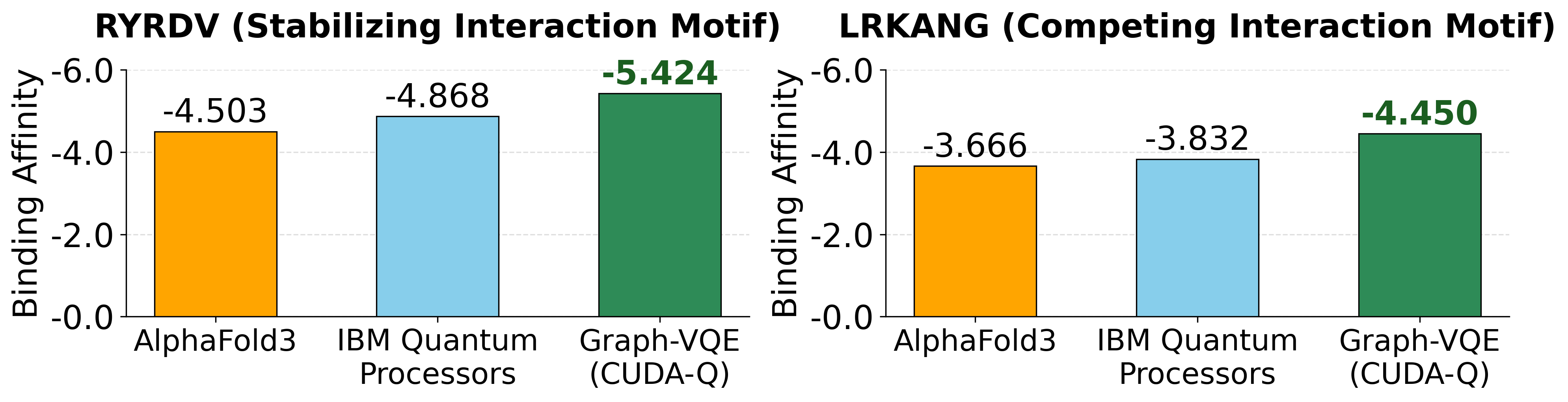}}
  \caption{Binding Affinity Comparison (kcal/mol, lower is better.)}
  \label{fig:affinity_comparison}
\end{figure}

Scaling VQE for large-scale biological systems requires shifting toward parallel optimization, but several architectural bottlenecks currently stand in the way. First, breaking a quantum problem into parallel parts is difficult; simple geometric partitioning often splits strongly connected qubits, creating errors that ruin the final result~\cite{fortunato2010community, djidjev2016graph}. Second, distributing these connected parts across separate processors without considering their relationship causes high communication delays. While methods like circuit cutting can distribute circuits across processors, they create a new problem by requiring exponential amounts of classical computing time to reassemble the data~\cite{eddins2022doubling, tang2022cutting, zhang2022variational}. Additionally, protein molecules have very uneven workloads: some areas, like hydrophobic clusters, have dense and strong interactions, while others are sparse and weak~\cite{miyazawa1985estimation, miyazawa1996residue}. Because the way a protein folds is directly tied to how its residues are sequenced and spaced, the physical structure of the protein defines the complexity of the math problem~\cite{plaxco1998contact, ivankov2003contact}. Fortunately, proteins naturally organize into ``modular communities'', suggesting that biologically-informed boundaries can be used to split the problem more effectively~\cite{vishveshwara2002protein, grant2019modular, bagler2007assortative}. To be truly efficient, a parallel architecture must identify these natural boundaries and use adaptive scheduling to ensure fast-solving sections do not sit idle while waiting for more complex parts to finish~\cite{ghale2017task}.

Although hybrid quantum-classical algorithms show promise, the VQE faces a critical ``trilemma'' of scalability, trainability, and measurement overhead~\cite{tilly2022variational}. Existing approaches address these axes in isolation: circuit simplification targets barren plateaus but cannot resolve landscape flattening caused by global cost functions and hardware noise~\cite{mcclean2018barren, cerezo2021cost, wang2021noise}; subspace partitioning methods rely on general-purpose graph algorithms that ignore molecular coupling structure~\cite{karypis1998multilevel, fortunato2010community}; and measurement optimization reduces sampling time without parallelizing parameter training~\cite{stein2022eqc}. No current framework combines Hamiltonian-aware partitioning with multi-QPU optimization-level parallelism without incurring prohibitive classical overhead.

To address these systemic bottlenecks, Graph-VQE is introduced as a Hamiltonian-aware optimization framework designed for scalable, multi-processor parallelization, building upon the existing research~\cite{zhang2025quantum}. The framework is built on NVIDIA's CUDA-Q platform~\cite{the_cuda_q_development_team_2026_19057431} for three reasons: (i) Current NISQ devices suffer from gate errors, decoherence, and limited circuit depth that prevent VQE from reaching chemical accuracy on molecules of practical size~\cite{preskill2018quantum, wang2021noise}. CUDA-Q eliminates hardware-induced errors such as gate infidelity, decoherence, and crosstalk, isolating algorithmic performance from device artifacts. Although finite-shot sampling still introduces statistical noise, this shot noise is well-characterized and reducible by increasing the number of measurements, unlike the systematic and unpredictable errors on current physical devices. (ii) Unlike standard CPU-based quantum simulators, CUDA-Q offers JIT-compiled quantum kernels, enabling exact state vector simulation~\cite{bayraktar2023cuquantum, brown2025multi}. Its multi-QPU simulation backend emulates multiple QPUs on a single machine, allowing task-parallel workflows that mirror the architecture of future multi-QPU systems. (iii) While CUDA-Q natively supports Hamiltonian term batching to parallelize energy measurement, it does not parallelize the optimization itself, since all GPUs still evaluate the same parameter vector. Graph-VQE extends CUDA-Q by adding optimization-level parallelism: the Hamiltonian interaction graph is partitioned into weakly coupled blocks, each block performs restricted updates on the same full-Hamiltonian objective while the remaining parameters are held fixed, and a short global refinement stage then updates all parameters jointly from the fused block solution. As shown in Figure~\ref{fig:affinity_comparison}, Graph-VQE on CUDA-Q achieves stronger binding affinities than both AlphaFold3 and IBM quantum processors across representative protein sequences. We summarize our architectural contributions as follows:

\begin{itemize}
    \item \textbf{Biologically-Informed Hamiltonian Partitioning:} We design a graph-based decomposition strategy that exploits amino acid sequence structures by leveraging Miyazawa-Jernigan contact potentials to construct a weighted interaction graph from the protein Hamiltonian. Using Louvain community detection, the framework identifies weakly-coupled qubit clusters that correspond to natural physical boundaries in the protein structure, enabling effective problem decomposition while preserving essential inter-residue correlations.

    \item \textbf{Hierarchical Multi-QPU Parallelization:} We develop a two-level parallel execution strategy that combines task parallelism for concurrent block optimization with data parallelism for global refinement. Multiple emulated QPUs simultaneously optimize independent parameter subsets, followed by Hamiltonian batching across GPUs during the full-parameter refinement stage, enabling multi-GPU scaling while preserving the method's optimization-quality gains.

    \item \textbf{Qiskit-CUDA-Q Adaptation:} We design a translation layer that bridges Qiskit's operator logic and optimization libraries with CUDA-Q's JIT-compiled quantum kernels and multi-QPU simulation backend. This integration enables domain scientists to leverage familiar Qiskit workflows while benefiting from GPU-accelerated state vector simulation and high-performance parallel execution.

\end{itemize}



\section{Background}
\label{sec:background}

\subsection{Protein Modeling and Energy Interpretation}
The protein folding problem aims to predict a protein's three-dimensional structure from its amino acid sequence, following the thermodynamic hypothesis that a protein naturally adopts the conformation at the global minimum of its free energy landscape~\cite{dill2012protein, chikenji2006shaping}. To make this search computationally tractable, lattice models restrict amino acids to discrete vertices on a grid. In this work, a tetrahedral lattice is used where the protein's shape is defined by a sequence of turn vectors between consecutive residues~\cite{robert2021resource}.

The total energy of a given conformation combines two components: physical interaction energies from the Miyazawa-Jernigan (MJ) statistical contact potential~\cite{miyazawa1996residue}, which quantifies attractive and repulsive forces between amino acid pairs based on a matrix of 210 unique pairwise energies, and geometric penalty terms that assign high energy costs to physically impossible configurations such as chirality violations, immediate back-turns, and lattice site overlaps~\cite{robert2021resource, fingerhuth2018quantum, kuroiwa2021penalty}. The calibration of penalty coefficients is important: values must be large enough to discourage invalid structures without distorting the optimization landscape~\cite{kuroiwa2021penalty}. The formal Hamiltonian construction and qubit encoding are presented in Section~\ref{sec:methodology}.

The final optimized energy produced by VQE serves as a validity indicator for the predicted structure. A negative total energy confirms that all geometric constraints were satisfied and that the conformation is dominated by stabilizing amino acid interactions, since MJ contact energies are defined as negative values~\cite{miyazawa1985estimation, robert2021resource}. Conversely, a positive total energy signals an invalid structure: the penalty weights are calibrated to overwhelm any physical stabilization, so even a single geometric violation produces a net positive energy~\cite{robert2021resource, kuroiwa2021penalty}.

\subsection{Applications of Protein Structure Prediction}
Predicting how a protein folds has transformative implications for medicine and biotechnology, as a protein's specific shape dictates its biological function. The primary application lies in computer-aided drug discovery, where knowing the precise structure of a target protein, such as a viral enzyme or a cancer-related receptor, allows researchers to design molecules that bind to it with high affinity, effectively locking the protein and neutralizing its harmful effects \cite{8585034}. Beyond therapeutics, structure prediction is essential for protein engineering, enabling the creation of synthetic enzymes tailored for industrial tasks like plastic degradation or carbon capture. By saving years of expensive and labor-intensive laboratory experiments like X-ray crystallography, computational modeling accelerates the identification of disease mechanisms and the development of personalized treatments, bridging the gap between genomic data and actionable medical solutions \cite{jumper2021highly, outeiral2021prospects}.

\subsection{NVIDIA CUDA-Q Platform}
\label{sec:cudaq}
CUDA-Q is an open-source platform for heterogeneous quantum-classical computing~\cite{the_cuda_q_development_team_2026_19057431}. Its programming model is built around \textit{quantum kernels}, parameterized circuit functions that are just-in-time (JIT) compiled and reused across optimization iterations without reconstruction overhead. CUDA-Q provides multiple simulation backends trading off qubit capacity, accuracy, and parallelism; of particular relevance is the multi-QPU backend, which treats each GPU as an independent virtual quantum processing unit for parallel circuit execution with native Hamiltonian term batching. However, CUDA-Q primarily targets molecular chemistry workflows with the UCCSD ansatz and does not provide direct integration with Qiskit, whose ecosystem includes hardware-efficient ansatz templates, operator algebra, and optimization libraries essential for non-chemistry quantum applications.
\section{Related Work}

A central challenge in Variational Quantum Eigensolver (VQE) is designing optimization strategies that converge reliably to low-energy solutions despite barren plateaus and local minima~\cite{mcclean2018barren, cerezo2021variational}. ADAPT-VQE~\cite{grimsley2019adaptive} addresses this by growing the ansatz iteratively rather than fixing its structure in advance. At each step, ADAPT-VQE evaluates the energy gradient with respect to a pool of candidate operators and appends the operator with the largest gradient to the circuit. This adaptive construction produces compact circuits tailored to the specific problem, avoiding the unnecessary depth of fixed hardware-efficient ansatzes~\cite{kandala2017hardware}. However, each operator selection step requires evaluating gradients for the entire pool, and the number of operators needed to reach convergence can grow substantially for larger systems, increasing the total number of optimization iterations. CVaR-VQE~\cite{barkoutsos2020improving} takes a different approach by modifying the cost function itself. Instead of minimizing the standard expectation value $\langle H \rangle$, CVaR-VQE minimizes the Conditional Value at Risk at confidence level $\alpha$, defined as the average energy of the lowest-$\alpha$ fraction of measurement outcomes. By focusing the optimizer on the low-energy tail of the distribution, CVaR-VQE biases the search toward promising regions of the landscape and reduces sensitivity to high-energy outliers caused by noise or poor initialization. The parameter $\alpha$ controls this trade-off: smaller values focus more aggressively on the best samples but increase variance, while larger values approach the standard expectation value. Other algorithmic improvements include subspace partitioning based on mutual information~\cite{zhang2022variational}, symmetry-preserving ansatzes~\cite{kirby2021contextual}, and penalty-based constraint encoding that prunes invalid configurations from the search space~\cite{fingerhuth2018quantum, kuroiwa2021penalty}.

On the scalability side, several strategies have been proposed to distribute VQE workloads across multiple processors. NVIDIA's CUDA-Q framework supports Hamiltonian term batching~\cite{the_cuda_q_development_team_2026_19057431}, which partitions the Hamiltonian's Pauli terms across GPUs so that each processor computes expectation values for a subset of terms before aggregating results. This achieves near-linear speedup for energy evaluation but parallelizes measurement rather than optimization: all GPUs evaluate the same parameter vector, and parameter updates remain sequential. Ensembled Quantum Computing (EQC)~\cite{stein2022eqc} takes a different approach by partitioning the quantum state space and running multiple smaller VQE instances in parallel, combining results through classical post-processing. EQC demonstrates that decomposing large problems into subproblems can achieve comparable accuracy with reduced per-circuit complexity. Other parallelization techniques include entanglement forging~\cite{eddins2022doubling} and circuit cutting~\cite{tang2022cutting}, which distribute circuits across QPUs but incur exponential classical overhead in the number of cuts. Graph-based partitioning has also proven effective in classical quantum molecular dynamics~\cite{djidjev2016graph, ghale2017task}, demonstrating that physics-informed decomposition outperforms naive geometric methods. Graph-theoretic analyses further reveal that proteins possess natural modular community structure~\cite{vishveshwara2002protein, bagler2007assortative}, suggesting that biologically informed boundaries can guide problem decomposition. However, existing parallelization approaches generally do not consider the Hamiltonian's interaction structure when partitioning, meaning strongly coupled qubits may be separated across subspaces, requiring expensive recombination to recover correlations.

\section{Methodology}
\label{sec:methodology}

Graph-VQE is a scalable hybrid quantum-classical optimization framework that accelerates the Variational Quantum Eigensolver (VQE) for large-scale molecular simulations. The framework combines graph-theoretic problem decomposition with a hierarchical multi-QPU parallelization strategy by extending CUDA-Q. The molecular Hamiltonian typically has a sparse structure where certain qubits interact more strongly than others. Graph-VQE exploits this structure by decomposing the high-dimensional optimization landscape into smaller subspaces that can be processed concurrently on distributed quantum resources.

\subsection{Problem Formulation and Hamiltonian Encoding}
The input to Graph-VQE is the primary amino acid sequence $S = \{s_1, s_2, \dots, s_L\}$, where $s_i$ represents the amino acid type at position $i$. The protein conformation is modeled on a 3D tetrahedral lattice using a coarse-grained approach, where each shape is described by a sequence of discrete turns rather than absolute coordinates. This turn-based representation reduces the degrees of freedom and enables efficient qubit encoding.

The total energy Hamiltonian $H_{total}$ governs the folding process, consisting of physical interactions and geometric penalty terms:
\begin{equation}
\label{eq:htotal}
    H_{total} = H_{interaction} + H_{penalty}
\end{equation}
The attractive forces that drive folding come from the Miyazawa-Jernigan (MJ) interaction model. This statistical potential assigns a contact energy $\epsilon(s_i, s_j)$ between amino acids $s_i$ and $s_j$ based on how frequently they appear in contact in known protein structures. The interaction term $H_{interaction}$ sums these energies over all non-covalent nearest-neighbor contacts in the lattice conformation:
\begin{equation}
\label{eq:hint}
    H_{interaction} = \sum_{i < j} \epsilon(s_i, s_j) \cdot \delta_{contact}(i, j)
\end{equation}
where $\delta_{contact}(i, j) = 1$ if residues $i$ and $j$ are non-adjacent in sequence but occupy neighboring lattice sites, and $0$ otherwise. The MJ potential matrix provides 210 unique pairwise energies for the 20 standard amino acids, with more negative values indicating stronger favorable interactions (e.g., hydrophobic-hydrophobic contacts).

To enforce the physical realism of the polypeptide chain, three distinct penalty terms are incorporated, each weighted by a coefficient $\lambda$:
\begin{equation}
\label{eq:hpenalty}
    H_{penalty} = \lambda_{chiral} H_{chiral} + \lambda_{back} H_{back} + \lambda_{overlap} H_{overlap}
\end{equation}
Here, $H_{chiral}$ enforces the correct tetrahedral chirality for the amino acid side chains, $H_{back}$ prevents the chain from reversing direction immediately onto itself (i.e., $s_{i+2}$ overlapping $s_i$), and $H_{overlap}$ penalizes non-local self-intersections where two distinct monomers occupy the same lattice site. All penalty parameters are set uniformly to $\lambda_{chiral} = \lambda_{back} = \lambda_{overlap} = 10.0$, creating a high energy barrier against invalid configurations. The goal of VQE is to find the parameter vector $\boldsymbol{\theta}^*$ that minimizes the expectation value of $H_{total}$:
\begin{equation}
\label{eq:vqe_objective}
    \boldsymbol{\theta}^* = \arg\min_{\boldsymbol{\theta}} E(\boldsymbol{\theta}) = \arg\min_{\boldsymbol{\theta}} \langle \psi(\boldsymbol{\theta}) | H_{total} | \psi(\boldsymbol{\theta}) \rangle
\end{equation}
where $|\psi(\boldsymbol{\theta})\rangle$ is the parameterized quantum state prepared by the variational ansatz.

\begin{figure}[tbp]
  \centering
  \resizebox{\linewidth}{!}{\includegraphics{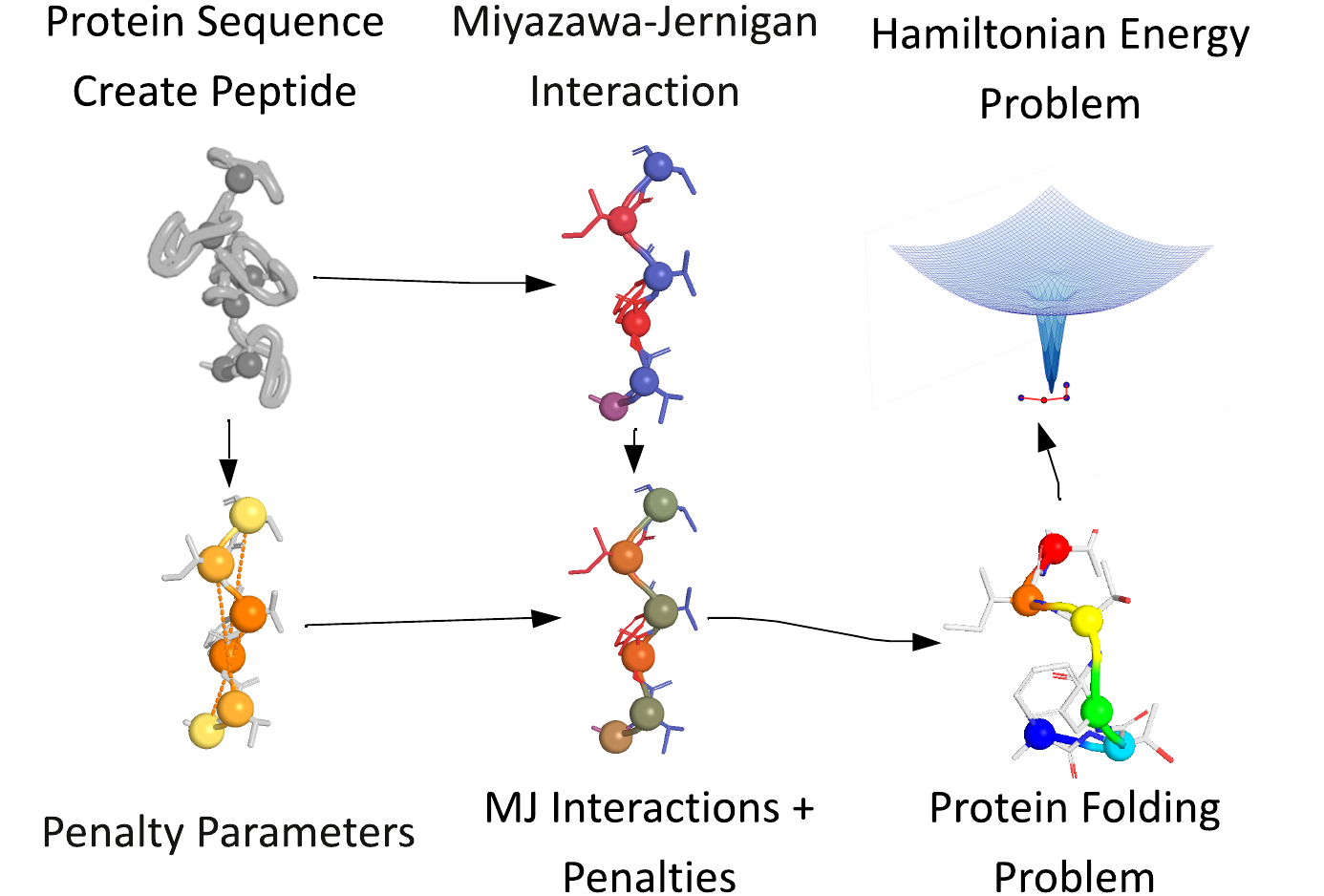}}
  \caption{Hamiltonian Construction Pipeline for Lattice Protein Folding}
  \label{fig:protein}
\end{figure}

The total Hamiltonian $H_{total}$ from Eq.~\ref{eq:htotal} is mapped to a qubit representation. For a protein of length $L$, encoding the turn sequence requires $N$ qubits:
\begin{equation}
\label{eq:nqubits}
    N = 2(L - 1)
\end{equation}
Each of the $L-1$ turns between consecutive residues is encoded using 2 qubits, representing four possible tetrahedral directions. The constrained optimization problem is then expressed as a qubit Hamiltonian in the form of a linear combination of Pauli strings:
\begin{equation}
\label{eq:pauli_ham}
    H = \sum_{k=1}^{M} c_k P_k, \quad P_k \in \{I, X, Y, Z\}^{\otimes N}
\end{equation}
Here, $c_k \in \mathbb{R}$ are real coefficients derived from the MJ interaction energies (Equation~\ref{eq:hint}) and penalty weights (Equation~\ref{eq:hpenalty}), and $M$ is the total number of Pauli terms. The set $\{I, X, Y, Z\}$ denotes the single-qubit Pauli matrices: the identity $I$ and the three Pauli operators $X$, $Y$, $Z$. The notation $\{I, X, Y, Z\}^{\otimes N}$ represents all possible $N$-qubit tensor products of these matrices. For protein folding problems, $H$ is typically sparse but contains high-weight non-local Pauli terms because long-range physical interactions arise when distant residues fold close together. Figure~\ref{fig:protein} illustrates the complete Hamiltonian construction pipeline from amino acid sequence to qubit Hamiltonian.

\subsection{Qiskit-CUDA-Q Adaptation}
To bridge the Qiskit interoperability gap identified in Section~\ref{sec:cudaq}, Graph-VQE proposes a custom adaptation layer that translates Qiskit-constructed Hamiltonians into CUDA-Q spin operators and maps the EfficientSU2 ansatz structure to JIT-compiled CUDA-Q kernels, enabling domain scientists to leverage familiar Qiskit workflows while benefiting from GPU-accelerated state vector simulation.

Graph-VQE leverages CUDA-Q's JIT-compiled kernels and multi-QPU simulation backend (Section~\ref{sec:cudaq}). While CUDA-Q natively uses this backend for data parallelism, distributing Hamiltonian terms across QPUs for parallel expectation value evaluation, Graph-VQE repurposes it for task parallelism, assigning different parameter subsets to different emulated QPUs so that multiple blocks are optimized simultaneously.

Graph-VQE thus uses the multi-QPU backend for two complementary purposes: during block optimization, each QPU updates a different parameter subset while the rest remain fixed, enabling concurrent block updates scored against the full Hamiltonian; during global refinement, the same QPUs switch to CUDA-Q's native Hamiltonian term batching to accelerate joint full-parameter evaluations.

\begin{figure}[tbp]
  \centering
  \resizebox{\linewidth}{!}{\includegraphics{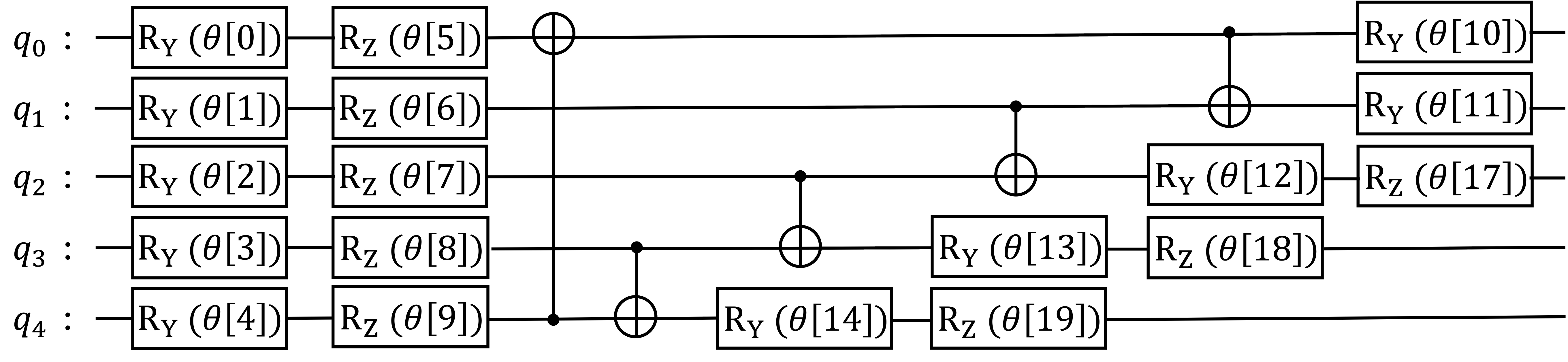}}
  \caption{Hardware-Efficient EfficientSU2 Ansatz Structure}
  \label{fig:efficientsu2}
\end{figure}

The variational ansatz follows a hardware-efficient EfficientSU2 design, as illustrated in Figure~\ref{fig:efficientsu2}. The circuit consists of $D$ repetitions (layers). Each layer applies parameterized single-qubit rotation gates to all $N$ qubits (from Eq.~\ref{eq:nqubits}), followed by an entangling layer of CNOT gates. The rotation gates are defined as:
\begin{align}
\label{eq:rotation_gates}
    R_y(\theta) &= e^{-i\frac{\theta}{2}Y} = \begin{pmatrix} \cos\frac{\theta}{2} & -\sin\frac{\theta}{2} \\ \sin\frac{\theta}{2} & \cos\frac{\theta}{2} \end{pmatrix} \\
    R_z(\theta) &= e^{-i\frac{\theta}{2}Z} = \begin{pmatrix} e^{-i\frac{\theta}{2}} & 0 \\ 0 & e^{i\frac{\theta}{2}} \end{pmatrix} \nonumber
\end{align}
where $\theta \in [-\pi, \pi]$ is the rotation angle, and $Y$, $Z$ are the Pauli matrices. The total number of variational parameters is:
\begin{equation}
\label{eq:nparams}
    P = 2N(D + 1)
\end{equation}
where the factor of 2 accounts for both $R_y$ and $R_z$ gates per qubit per layer, and $D+1$ includes the initial rotation layer plus $D$ repeated layers. The entanglement pattern can be configured as linear (nearest-neighbor), circular (linear with wrap-around), or full (all-to-all), allowing the ansatz to balance expressibility against circuit depth based on the problem requirements. Using the CUDA-Q kernel builder API described above, the ansatz is compiled into a parameterized kernel $K(\boldsymbol{\theta})$ that accepts a flattened parameter vector $\boldsymbol{\theta} \in \mathbb{R}^P$ and implements the unitary transformation $U(\boldsymbol{\theta})$ preparing $|\psi(\boldsymbol{\theta})\rangle = U(\boldsymbol{\theta})|0\rangle^{\otimes N}$. Because the kernel is JIT-compiled once and reused across all optimization iterations, parameter updates from the classical optimizer are injected with minimal latency.

\subsection{Biologically-Informed Hamiltonian Partitioning}
\label{sec:subspace_decomp}
The standard VQE approach optimizes the entire parameter vector $\boldsymbol{\theta}$ simultaneously, but suffers from barren plateaus and high classical optimization overhead ($O(P^2)$ or $O(P^3)$) as $P$ increases. To address this, Graph-VQE introduces a Hamiltonian-aware partitioning strategy that clusters qubits based on their interaction strength. A weighted interaction graph $G = (V, E, W)$ is constructed from the Pauli Hamiltonian (Equation~\ref{eq:pauli_ham}). The vertex set $V = \{q_0, q_1, \dots, q_{N-1}\}$ represents the $N$ system qubits, and an edge $e_{ij} \in E$ exists between qubits $q_i$ and $q_j$ if they appear simultaneously in any non-identity Pauli term $P_k$. The edge weight $w_{ij}$ quantifies the interaction strength between qubits $q_i$ and $q_j$ by aggregating the absolute coefficients:
\begin{equation}
\label{eq:edge_weight}
    w_{ij} = \sum_{k: P_k^{(i)}, P_k^{(j)} \neq I} |c_k|
\end{equation}
where $P_k^{(i)}$ denotes the Pauli operator acting on qubit $i$ in term $k$, and $c_k$ are the coefficients from Eq.~\ref{eq:pauli_ham}. This graph captures the entanglement requirements imposed by the problem Hamiltonian: strongly coupled qubits (high $w_{ij}$) should be optimized together, while weakly coupled qubits can be optimized semi-independently.

Crucially, this interaction graph directly exploits the amino acid sequence structure through the qubit encoding. Recall from Eq.~\ref{eq:nqubits} that each consecutive pair of qubits $(q_{2i}, q_{2i+1})$ encodes the turn between residues $s_i$ and $s_{i+1}$ in the sequence. The Hamiltonian coefficients $c_k$ are derived from Miyazawa-Jernigan contact energies $\epsilon(s_i, s_j)$, which depend on the specific amino acid types at positions $i$ and $j$. Therefore, the edge weights $w_{ij}$ in the interaction graph inherit the biological properties of the amino acid sequence: hydrophobic residues that tend to cluster together produce strong edge weights between their corresponding qubits, while polar or charged residues at the protein surface yield weaker connections. By partitioning this graph, Graph-VQE groups qubits whose associated residues interact strongly according to the MJ potential, effectively decomposing the optimization problem along natural biochemical boundaries defined by the input sequence.

\begin{figure}[tbp]
  \centering
  \resizebox{\linewidth}{!}{\includegraphics{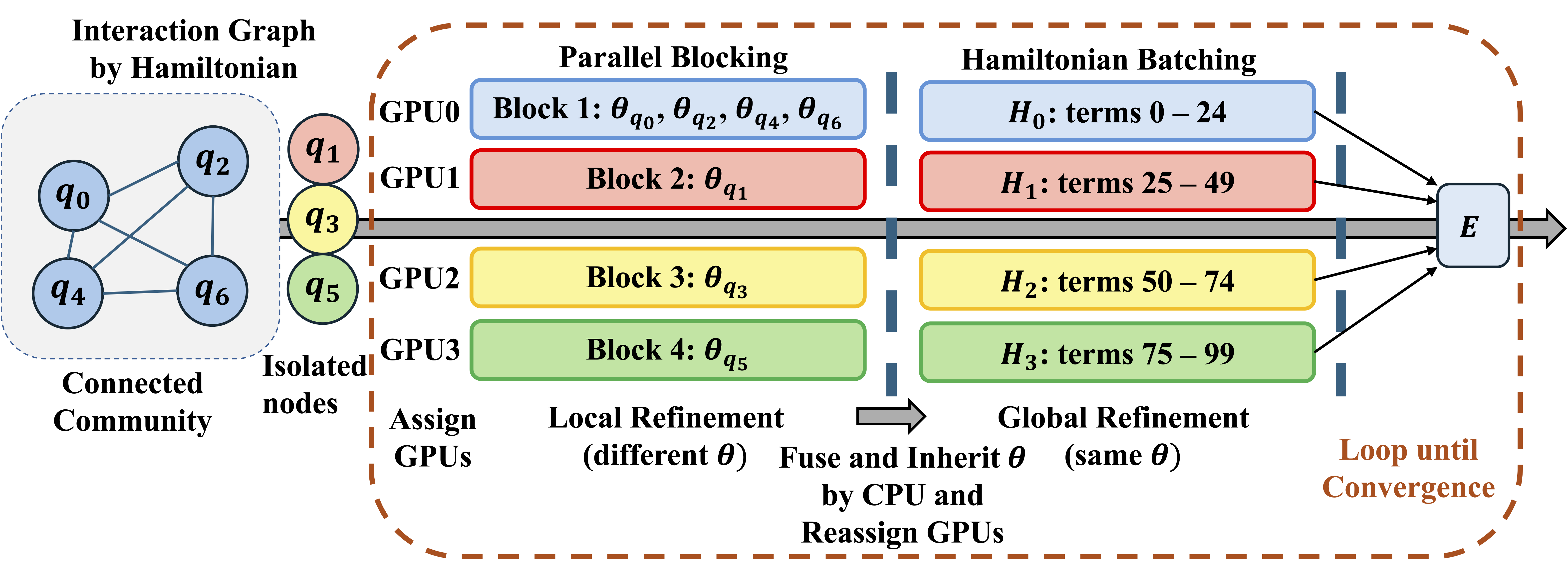}}
  \caption{Graph-VQE Optimization Workflow. A weighted qubit interaction graph is built from the Hamiltonian coefficients and partitioned via Louvain community detection into blocks of strongly coupled qubits. Each block is dispatched to a separate GPU for concurrent low-dimensional VQE optimization while all other parameters remain frozen, using the same full Hamiltonian as the objective. A final global refinement phase then jointly updates all parameters with GPU-parallel Hamiltonian batching to recover cross-block correlations. The variational parameters are stored and updated on the CPU; GPUs accelerate circuit evaluations when computing expectation values.}
  \label{fig:optimization}
\end{figure}

The graph $G$ is then partitioned into $K$ disjoint communities (blocks) $\{B_1, B_2, \dots, B_K\}$ using the Louvain community detection algorithm. The algorithm determines $K$ automatically, and each block $B_i \subseteq V$ is a subset of qubits satisfying $\bigcup_{i=1}^{K} B_i = V$ and $B_i \cap B_j = \emptyset$ for $i \neq j$. The Louvain algorithm maximizes the modularity $Q$ of the partition, defined as:
\begin{equation}
\label{eq:modularity}
    Q = \frac{1}{2m} \sum_{i,j} \left[ w_{ij} - \frac{k_i k_j}{2m} \right] \delta(c_i, c_j)
\end{equation}
where $m = \frac{1}{2}\sum_{i,j} w_{ij}$ is the total edge weight, $k_i = \sum_j w_{ij}$ is the weighted degree of node $i$, $c_i$ is the community assignment of node $i$, and $\delta$ is the Kronecker delta. High modularity indicates dense intra-community connections relative to a random graph, ensuring that strongly interacting qubits are grouped together. This partitioning keeps the most strongly coupled variables inside the same block, reducing the amount of cross-block coordination that must later be handled by joint global updates. Blocks smaller than a configurable threshold $S_{\min}$ are iteratively merged with neighbors to prevent fragmentation. Figure~\ref{fig:optimization} illustrates the complete Graph-VQE optimization workflow from graph construction through parallel block optimization to global refinement.

\subsection{Hierarchical Multi-QPU Parallelization}
In the first phase, the $P$ parameters (from Eq.~\ref{eq:nparams}) are partitioned into $K$ subsets $\{\boldsymbol{\theta}_{B_1}, \boldsymbol{\theta}_{B_2}, \dots, \boldsymbol{\theta}_{B_K}\}$ corresponding to the qubit blocks identified by modularity optimization (Eq.~\ref{eq:modularity}). Here, $\boldsymbol{\theta}_{B_i} \subset \boldsymbol{\theta}$ denotes the subset of variational parameters associated with qubits in block $B_i$. Because only the coordinates in $\boldsymbol{\theta}_{B_i}$ are updated while all other parameters are held fixed, each block subproblem is lower-dimensional and the blocks can be processed concurrently. A worker pool of $N_{GPU}$ available GPUs handles the parallel execution. The blocks are distributed round-robin to the workers: worker $W_j$ (for $j \in \{0, 1, \dots, N_{GPU}-1\}$) processes blocks $\{B_i : i \mod N_{GPU} = j\}$. Each worker executes a local VQE loop for its assigned block on a dedicated QPU, minimizing the restricted global cost function:
\begin{equation}
\label{eq:block_energy}
    E_{B_i}(\boldsymbol{\theta}_{B_i}) = \langle \psi(\boldsymbol{\theta}) | H | \psi(\boldsymbol{\theta}) \rangle \big|_{\boldsymbol{\theta}_{\setminus B_i} = \text{const}}
\end{equation}
where $\boldsymbol{\theta}_{\setminus B_i} = \boldsymbol{\theta} \setminus \boldsymbol{\theta}_{B_i}$ denotes all parameters outside block $B_i$, held fixed during local optimization. Equation~\ref{eq:block_energy} is therefore a restriction of the same full-Hamiltonian objective, not a separate block Hamiltonian. The computational savings come from optimizing only $P_{B_i}$ active parameters at a time and from running different blocks concurrently. At most $\min(K, N_{GPU})$ blocks can be optimized simultaneously in this phase.

In the second phase, a global refinement step optimizes all $P$ parameters simultaneously to reintroduce joint updates across blocks. In the implementation, this step is a short SciPy optimization loop initialized from the fused block solution. Depending on the chosen optimizer, the loop uses either COBYLA or L-BFGS-B on the full parameter vector, and every objective evaluation still uses the full Hamiltonian. To distribute the measurement overhead, Graph-VQE uses Hamiltonian Batching: the $M$ terms are partitioned into $N_{GPU}$ subsets $\{H_1, H_2, \dots, H_{N_{GPU}}\}$, where each sub-Hamiltonian $H_j = \sum_{k \in \mathcal{I}_j} c_k P_k$ contains approximately $\lceil M / N_{GPU} \rceil$ Pauli terms. Each QPU evaluates the expectation value of its assigned sub-Hamiltonian in parallel, and the total energy is reconstructed via reduction:
\begin{equation}
\label{eq:ham_batching}
    E(\boldsymbol{\theta}) = \sum_{j=1}^{N_{GPU}} \langle \psi(\boldsymbol{\theta}) | H_j | \psi(\boldsymbol{\theta}) \rangle
\end{equation}
This data-parallel approach reduces the cost of each full-Hamiltonian evaluation inside the refinement loop. Hamiltonian batching accelerates the evaluation routine itself; it does not alter the global objective being optimized. Early stopping terminates the refinement loop once recent improvements remain below the prescribed tolerance.

Graph-VQE supports both derivative-free (COBYLA) and gradient-based (L-BFGS-B) optimization. For gradient-based optimization, analytical gradients are computed using the parameter-shift rule. For a rotation gate $R(\theta_i) = e^{-i\frac{\theta_i}{2}G}$ where $G$ is a generator with eigenvalues $\pm 1$ (such as Pauli matrices), the gradient of the expectation value is:
\begin{equation}
\label{eq:param_shift}
    \frac{\partial E(\boldsymbol{\theta})}{\partial \theta_i} = \frac{1}{2}\left[ E\left(\boldsymbol{\theta} + \frac{\pi}{2}\mathbf{e}_i\right) - E\left(\boldsymbol{\theta} - \frac{\pi}{2}\mathbf{e}_i\right) \right]
\end{equation}
where $\mathbf{e}_i$ is the unit vector in the $i$-th direction. Computing the full gradient requires $2P$ circuit evaluations. To minimize overhead, all shifted parameter evaluations are batched into parallel GPU executions. For block-level optimization, gradients are computed only for the $P_{B_i} = |\boldsymbol{\theta}_{B_i}|$ parameters in block $B_i$, reducing the cost from $2P$ to $2P_{B_i}$ evaluations per gradient step. The L-BFGS-B optimizer uses these gradients with a function tolerance $f_{tol}$ for energy convergence and gradient tolerance $g_{tol} = 0.1 \cdot f_{tol}$ for gradient norm convergence.

\subsection{Convergence Control}
The Graph-VQE algorithm proceeds each epoch with four steps: (1) generate qubit blocks using the Hamiltonian interaction graph (or geometric alternating offsets for comparison); (2) optimize blocks concurrently on available QPUs with block-level early stopping; (3) perform a short full-parameter optimization run using distributed Hamiltonian batching to jointly update all parameters; and (4) evaluate the epoch-level improvement against a patience threshold.

A three-tier early stopping mechanism prevents over-optimization and minimizes unnecessary quantum resource usage. At the block level, energy improvements are monitored within a sliding window of $W$ iterations. Let $\{E_1, E_2, \dots, E_n\}$ be the energy history for block $B_i$. The maximum recent improvement is:
\begin{equation}
\label{eq:block_converge}
    \Delta_{\max} = \max_{t \in [n-W, n]} |E_t - E_{t-1}|
\end{equation}
If $\Delta_{\max} < \epsilon_{tol}$ for $p_{block}$ consecutive checks, block optimization terminates early. The same window-based criterion applies to the global refinement phase with patience $p_{global}$, preventing excessive iterations once the joint full-parameter updates have plateaued. At the epoch level, the historical best energy $E^*_{best}$ is tracked. If no new best is discovered for $p_{epoch}$ consecutive epochs, training terminates. To initialize, parameters are warm-started from $\boldsymbol{\theta}^*_{best}$ rather than the previous epoch's final state. This ensures that optimization always proceeds from the best-known configuration.

\subsection{Result Interpretation and Structure Reconstruction}
\begin{figure}[tbp]
  \centering
  \resizebox{\linewidth}{!}{\includegraphics{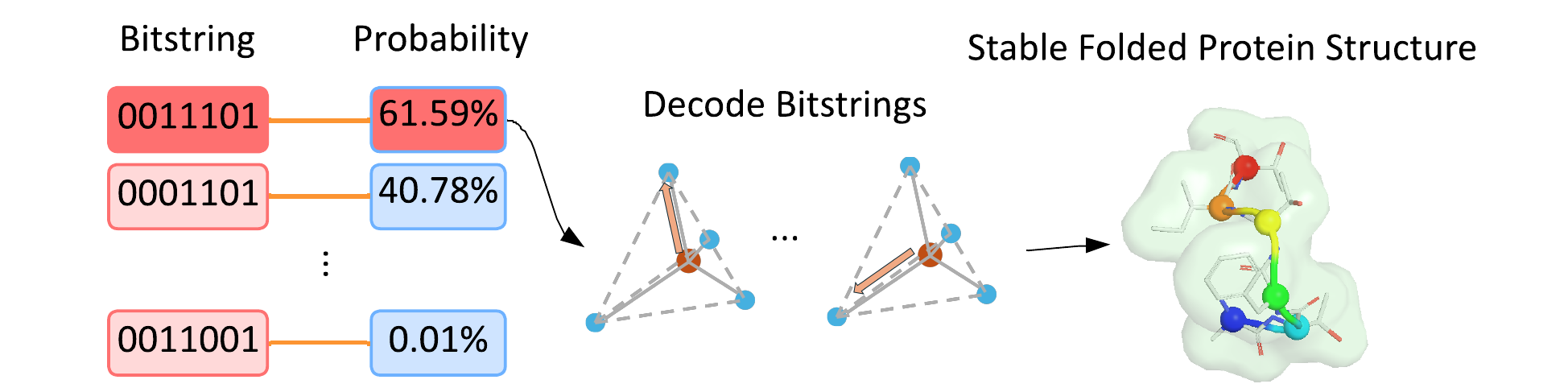}}
  \caption{Bitstring Decoding and 3D Structure Reconstruction}
  \label{fig:bitstring}
\end{figure}

Upon convergence of the VQE algorithm, the optimal parameter vector $\boldsymbol{\theta}^*$ is used to prepare the final quantum state $|\psi(\boldsymbol{\theta}^*)\rangle$. This state is sampled to obtain a probability distribution over the computational basis states (bitstrings). The bitstring with the highest probability is identified as the optimal turn sequence for the protein. This binary turn sequence is then decoded into a sequence of relative spatial moves on the tetrahedral lattice, as illustrated in Figure~\ref{fig:bitstring}. Each 2-bit segment of the bitstring encodes one of four possible turn directions on the tetrahedral lattice. Starting from the N-terminus fixed at the origin, these turns are sequentially applied to compute the 3D Cartesian coordinates of each amino acid residue. The cumulative displacement vectors trace out the backbone of the folded protein. The final output is an XYZ structure file representing the folded protein conformation, which corresponds to the global minimum of the energy landscape found by the Graph-VQE optimizer.

\section{Experiment}
\label{sec:experiments}

\subsection{Experimental Design}
This paper evaluates Graph-VQE on ten protein folding instances using the NVIDIA CUDA-Q multi-QPU simulation backend. The main accuracy comparisons use 6 NVIDIA L40s (48GB) GPUs, while a separate scaling study varies the GPU count. 
We evaluate Graph-VQE on two sets of peptide fragments that reflect distinct interaction regimes in biomolecular systems. Peptide Set 1 captures \textit{stabilizing interaction motifs} (e.g., 1fkf, 4mo4), where residue compositions favor relatively consistent hydrophobic packing and electrostatic stabilization. In contrast, Peptide Set 2 captures \textit{competing interaction motifs} (e.g., 3ibi, 6czf), where mixtures of aromatic, hydrophobic, and charged residues introduce competing forces and more rugged energy landscapes. This distinction enables a focused systems-level evaluation of whether Graph-VQE can robustly handle both stable and interaction-driven conformational regimes, rather than overfitting to a narrow or less challenging subset of instances. The evaluated sequences are:
\begin{itemize}
\item Peptide Set 1 (stabilizing interaction motifs): KPFKF (1fkf), NIGGF (4mo4), RYRDV (3eax), VKDRS (3ckz).
\item Peptide Set 2 (competing interaction motifs): ATFTIT (2v25), DGPHGM (1e2k), GIKAVM (3s0b), IQFHFH (3ibi), LRKANG (6czf), SIHSYS (1hdq).
\end{itemize}
Unless otherwise specified, all methods use the EfficientSU2 ansatz~\cite{kandala2017hardware} and 8192 measurement shots, and the ansatz depth (reps) is set to 4. For early stopping, we use an energy tolerance of $10^{-4}$, patience of 12 iterations, and a smoothing window of 40 iterations. The protein folding Hamiltonian uses MJ interaction parameters with penalty coefficients $(10, 10, 10)$ for chirality, backbone overlap, and side-chain overlap constraints. All experiments use a fixed random seed of 0 for reproducibility. We compare Graph-VQE against five existing optimization strategies:

\begin{itemize}
    \item \textbf{SeqVQE~\cite{zhang2025quantum}:} Standard single-GPU VQE execution using the gradient-free COBYLA optimizer. We evaluate it across six entanglement strategies: Circular, Full, Linear, Reverse Linear, Pairwise, and SCA.
    \item \textbf{HBatch~\cite{the_cuda_q_development_team_2026_19057431}:} A parallelized VQE baseline that distributes Hamiltonian terms across 4 GPUs to accelerate energy evaluation, also using the COBYLA optimizer.
    \item \textbf{EQC~\cite{stein2022eqc}:} Following the original implementation, EQC uses an SGD optimizer (learning rate 0.1, momentum 0) with a shallow ansatz (reps=1) for up to 250 epochs, employing asynchronous gradient updates across 4 GPUs where the first GPU to complete determines the parameter update.
    \item \textbf{CVaR-VQE~\cite{barkoutsos2020improving}:} Uses the Conditional Value at Risk (CVaR) objective with $\alpha=0.2$ to focus on the lower tail of the energy distribution. It employs single-GPU execution with the gradient-free COBYLA optimizer.
    \item \textbf{ADAPT-VQE~\cite{grimsley2019adaptive}:} An adaptive single-GPU strategy that iteratively grows the ansatz from a minimal operator pool (max 50 operators), starting from a zero initial state. It uses gradient threshold $10^{-5}$ and eigenvalue threshold $10^{-5}$ for operator selection, with COBYLA for parameter optimization.
    \item \textbf{Graph-VQE no\_Global (Ours, Ablation):} An ablation variant that performs only the Hamiltonian-aware block-wise optimization step and omits the subsequent global refinement phase. It is included to isolate and showcase the contribution of the global refinement phase within our framework.
    \item \textbf{Graph-VQE (Ours):} Uses Hamiltonian-aware graph partitioning for block-wise optimization with global refinement across 4 GPUs. It employs the L-BFGS-B gradient optimizer with epoch patience of 3 for efficient convergence.
\end{itemize}

\subsection{Main Analysis}
\begin{figure}[htbp]
  \centering
  \resizebox{\linewidth}{!}{\includegraphics{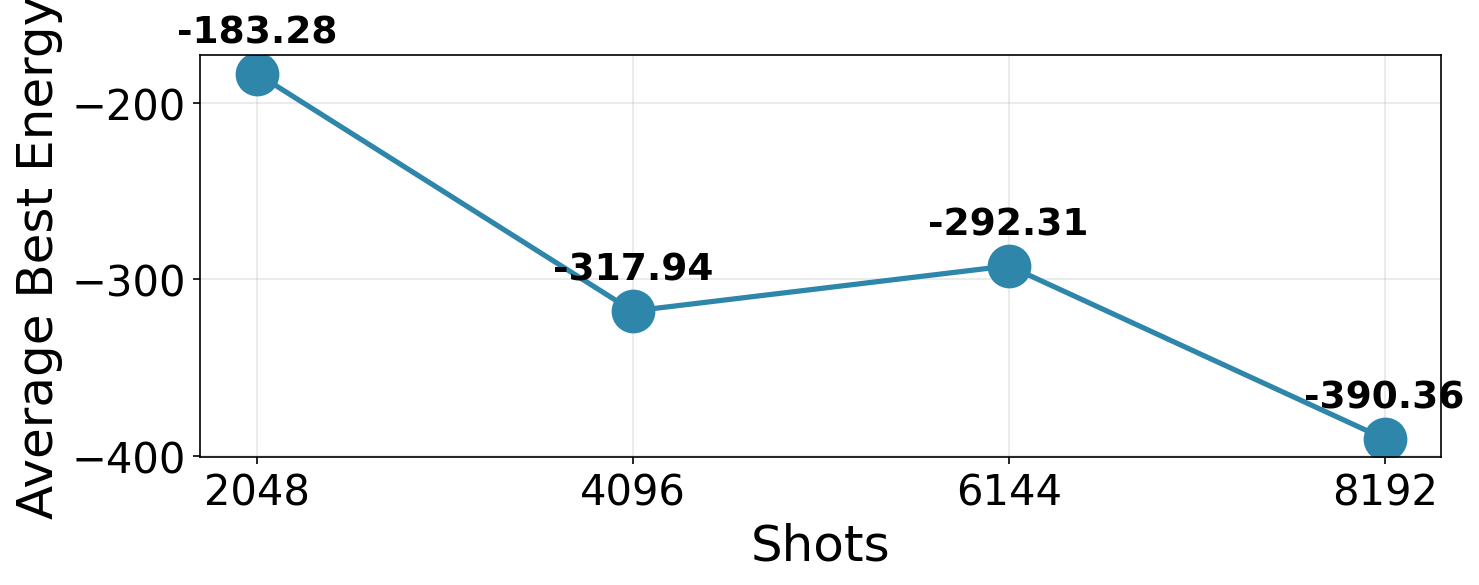}}
  \caption{Average best energy vs.\ measurement shots for Graph-VQE (Circular Entanglement), averaged across all protein sequences. Lower energy ($\downarrow$) indicates better optimization quality.}
  \label{fig:shots_energy}
\end{figure}

The choice of 8192 measurement shots is informed by a sensitivity study shown in Figure~\ref{fig:shots_energy}. We evaluate Graph-VQE under circular entanglement across all protein sequences at four shot counts: 2048, 4096, 6144, and 8192. The average best energy improves monotonically from $-183.28$ at 2048 shots to $-390.36$ at 8192 shots, demonstrating that higher shot budgets yield substantially better optimization quality by reducing the variance of expectation value estimates. Because VQE relies on finite-sample energy evaluations to guide the classical optimizer, insufficient shots introduce stochastic noise that can mislead gradient computations and cause premature convergence to suboptimal minima. At 8192 shots, the estimator variance scales as $O(1/\sqrt{8192})$, providing a practical balance between statistical accuracy and computational cost per iteration. This aligns with the broader quantum computing literature, where near-term applications typically adopt shot counts in the $2^{13}$--$2^{14}$ range to ensure reliable energy resolution~\cite{kandala2017hardware, peruzzo2014variational}.

\begin{figure}[htbp]
  \centering
  \resizebox{\linewidth}{!}{\includegraphics{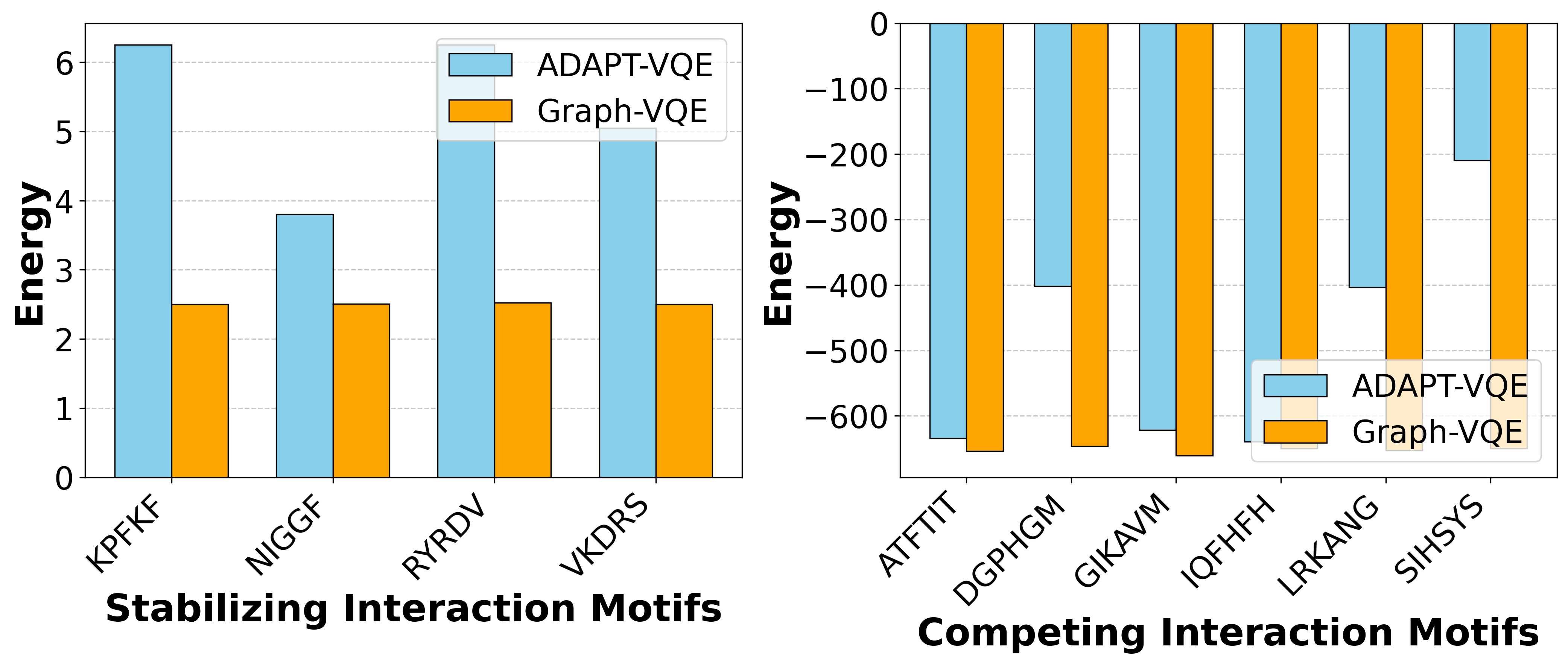}}
  \caption{ADAPT-VQE vs Graph-VQE (Circular Entanglement). Lower energy ($\downarrow$) indicates better optimization.}
  \label{fig:adapt_vs_graphvqe}
\end{figure}

Table~\ref{tab:vqe_by_entanglement} demonstrates that Graph-VQE achieves the lowest energy in 54 out of 60 test cases against the four fixed-ansatz baselines, with the full method dominating in 45 cases and the no\_Global ablation accounting for the remaining 9. The advantage is most pronounced for hexapeptide instances: under circular entanglement, Graph-VQE reaches energies of $-646$ to $-661$ versus $-162$ to $-256$ for the best baselines, a $2.5\times$--$4\times$ improvement. Graph-VQE also consistently outperforms ADAPT-VQE (Figure~\ref{fig:adapt_vs_graphvqe}), reaching below $-640$ on hexapeptides where ADAPT-VQE achieves only $-200$ to $-400$, showing that Hamiltonian-aware partitioning via Louvain community detection (Eq.~\ref{eq:edge_weight}) is more effective than adaptive ansatz construction. The 6 losses are all to EQC on pentapeptide (stabilizing interaction) instances, with pairwise entanglement accounting for 4 of the 6 (Table~\ref{tab:vqe_by_entanglement}d): these small systems yield only 3--4 Louvain blocks, limiting decomposition benefit, while EQC's shallow ansatz (reps$=$1) with SGD is better matched to their simpler landscapes, and pairwise entanglement's dense qubit-pair connectivity further reduces the modularity available for effective partitioning. The ablation reveals that no\_Global outperforms the full method in 9 of 60 cases, concentrated under reverse linear (4 cases) and SCA (3 cases) entanglement, where the L-BFGS-B global refinement converges to worse local minima than the block-only solution, indicating that certain entanglement topologies create landscapes where joint full-parameter updates disrupt favorable block-level configurations. Circular entanglement yields the most consistent advantage (10/10 wins), suggesting its nearest-neighbor connectivity with wrap-around aligns naturally with the Louvain-detected block boundaries.

\begin{figure}[htbp]
  \centering
  \resizebox{\linewidth}{!}{\includegraphics{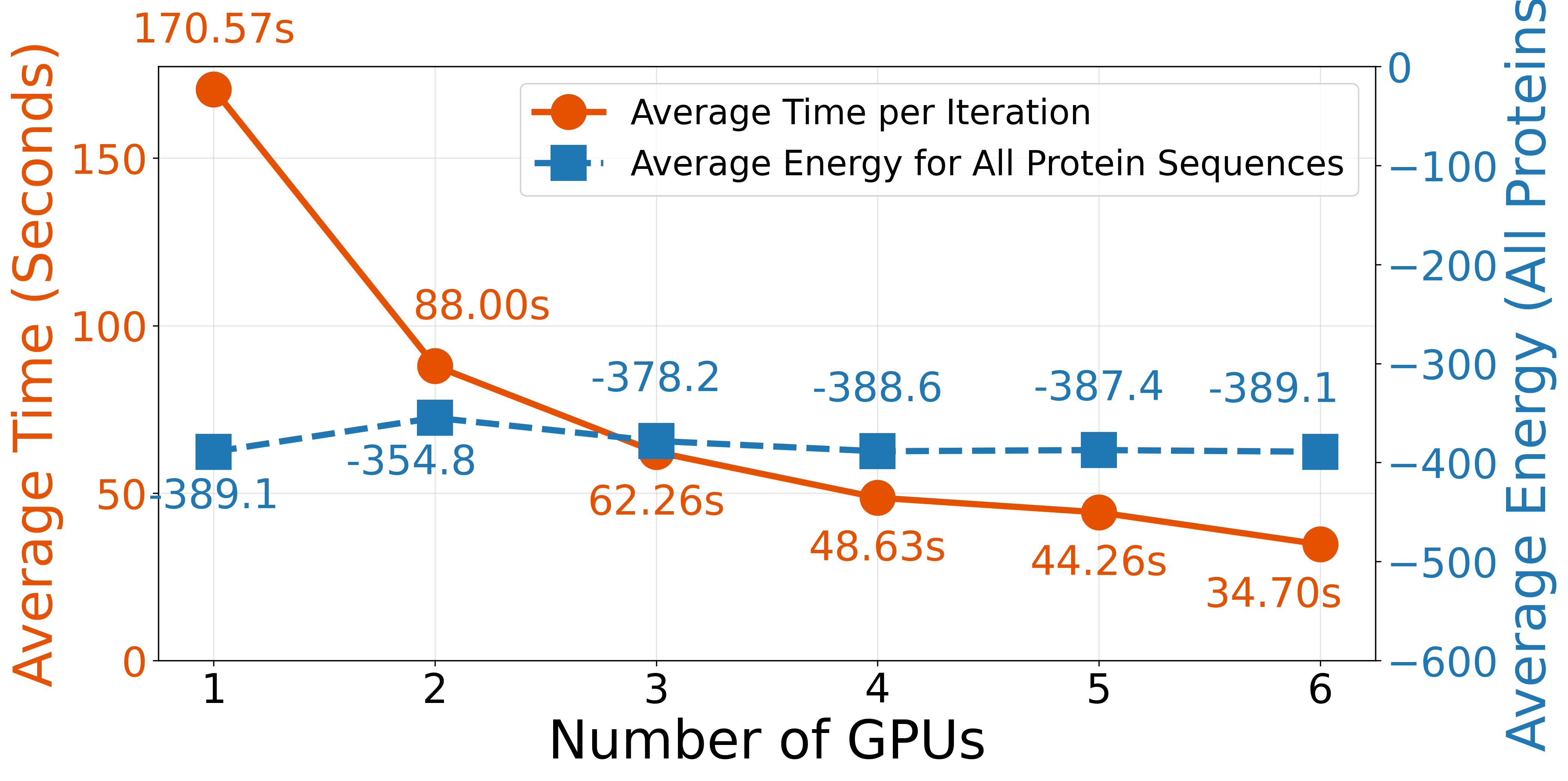}}
  \caption{GPU Scaling and Accuracy Preservation for Graph-VQE (Circular Entanglement).}
  \label{fig:gpu_scaling}
\end{figure}

Figure~\ref{fig:gpu_scaling} reports a within-method scaling study showing that Graph-VQE accelerates with additional GPUs while maintaining accuracy. Average time per iteration decreases monotonically from 170.57s (1~GPU) to 48.63s (4~GPUs) and 34.70s (6~GPUs), while the average final energy remains stable ($-354.8$ to $-389.1$) across all configurations. This confirms that the parallel block dispatch and Hamiltonian batching scale without degrading solution quality.

\begin{table*}[htbp]
  \caption{VQE Results by Entanglement Type}
  \label{tab:vqe_by_entanglement}
  \centering
  \footnotesize

  \begin{minipage}[t]{0.48\textwidth}
    \centering
    \textbf{(a) Circular Entanglement}
    \vspace{2mm}

    \resizebox{\linewidth}{!}{
    \begin{tabular}{ll| cccc|cc}
      \toprule
        \multicolumn{2}{c|}{{Protein}} & \multicolumn{4}{c|}{{Baselines}} & \multicolumn{2}{c}{\textbf{Ours}} \\
   \cmidrule(lr){1-2} \cmidrule(lr){3-6} \cmidrule(lr){7-8}
      Motifs & \makecell{Protein\\Sequences} & \makecell{SeqVQE} & \makecell{HBatch} & \makecell{EQC} & \makecell{CVaR-\\VQE} & \makecell{Graph-VQE\\no\_Global} & \makecell{\textbf{Graph-}\\\textbf{VQE}} \\
      \midrule
      \multirow{4}{*}{SI} & KPFKF & 7.4045 & 9.0541 & 3.7466 & 10.8751 & 3.3368 & \textbf{2.5032}\\
       & NIGGF & 7.5876 & 5.7048 & 3.7447 & 11.0278 & 2.5241 & \textbf{2.5037}\\
       & RYRDV & 7.1573 & 3.4109 & 3.7431 & 11.7058 & 2.5211 & \textbf{2.5209}\\
       & VKDRS & 7.0674 & 10.2042 & 3.7433 & 10.7095 & 2.5063 & \textbf{2.4983}\\
      \midrule
      \multirow{6}{*}{CI} & ATFTIT & -192.4622 & -173.8535 & -144.5634 & 247.0188 & -651.0444 & \textbf{-653.7233}\\
       & DGPHGM & -160.7019 & -208.9311 & -102.3001 & 282.4386 & -545.9451 & \textbf{-646.6160}\\
       & GIKAVM & -256.0308 & -207.7621 & -92.7553 & 270.8749 & -650.2604 & \textbf{-660.8168}\\
       & IQFHFH & -210.5125 & -225.1845 & -43.8601 & 244.6752 & -536.3909 & \textbf{-649.7793}\\
       & LRKANG & -161.8207 & -151.9016 & -25.2135 & 274.4735 & -222.4036 & \textbf{-652.8972}\\
       & SIHSYS & -189.9463 & -154.5095 & -118.1877 & 287.8844 & -641.8359 & \textbf{-649.7644}\\
      \bottomrule
    \end{tabular}}
  \end{minipage}
  \hfill
  \begin{minipage}[t]{0.48\textwidth}
    \centering
    \textbf{(b) Full Entanglement}
    \vspace{2mm}

    \resizebox{\linewidth}{!}{
    \begin{tabular}{ll| cccc|cc}
      \toprule
        \multicolumn{2}{c|}{{Protein}} & \multicolumn{4}{c|}{{Baselines}} & \multicolumn{2}{c}{\textbf{Ours}} \\
   \cmidrule(lr){1-2} \cmidrule(lr){3-6} \cmidrule(lr){7-8}
      Motifs & \makecell{Protein\\Sequences} & \makecell{SeqVQE} & \makecell{HBatch} & \makecell{EQC} & \makecell{CVaR-\\VQE} & \makecell{Graph-VQE\\no\_Global} & \makecell{\textbf{Graph-}\\\textbf{VQE}} \\
      \midrule
      \multirow{4}{*}{SI} & KPFKF & 7.1783 & 7.6228 & 3.7440 & 11.2154 & 2.5278 & \textbf{2.5006}\\
       & NIGGF & 7.7484 & 10.3383 & 3.7479 & 10.3284 & 3.5092 & \textbf{2.5093}\\
       & RYRDV & 9.9748 & 5.4703 & \textbf{3.7471} & 10.7324 & 4.4371 & 4.3237\\
       & VKDRS & 6.5686 & 10.1549 & 3.7453 & 10.5490 & 4.3373 & \textbf{3.0702}\\
      \midrule
      \multirow{6}{*}{CI} & ATFTIT & -261.4515 & -291.5842 & -343.5813 & 245.4055 & -435.4296 & \textbf{-658.6687}\\
       & DGPHGM & -276.3506 & -231.7292 & -171.9751 & 263.1004 & -642.3837 & \textbf{-654.9229}\\
       & GIKAVM & -235.1468 & -315.6424 & -219.4747 & 237.1520 & -108.0517 & \textbf{-429.1023}\\
       & IQFHFH & -270.8111 & -198.5136 & -253.4734 & 258.3943 & -318.5513 & \textbf{-542.8759}\\
       & LRKANG & -181.7336 & -179.8182 & -239.8358 & 269.1989 & \textbf{-647.7438} & -418.5219\\
       & SIHSYS & -286.2605 & -253.8739 & -274.3504 & 222.4124 & \textbf{-536.1251} & -394.5061\\
      \bottomrule
    \end{tabular}}
  \end{minipage}

  \vspace{4mm}

  \begin{minipage}[t]{0.48\textwidth}
    \centering
    \textbf{(c) Linear Entanglement}
    \vspace{2mm}

    \resizebox{\linewidth}{!}{
    \begin{tabular}{ll| cccc|cc}
      \toprule
        \multicolumn{2}{c|}{{Protein}} & \multicolumn{4}{c|}{{Baselines}} & \multicolumn{2}{c}{\textbf{Ours}} \\
   \cmidrule(lr){1-2} \cmidrule(lr){3-6} \cmidrule(lr){7-8}
      Motifs & \makecell{Protein\\Sequences} & \makecell{SeqVQE} & \makecell{HBatch} & \makecell{EQC} & \makecell{CVaR-\\VQE} & \makecell{Graph-VQE\\no\_Global} & \makecell{\textbf{Graph-}\\\textbf{VQE}} \\
      \midrule
      \multirow{4}{*}{SI} & KPFKF & 6.9258 & 8.3821 & \textbf{3.5880} & 10.9563 & 3.6743 & 3.7520\\
       & NIGGF & 6.3623 & 7.4071 & 3.6443 & 10.8705 & 4.0343 & \textbf{2.5116}\\
       & RYRDV & 6.0754 & 6.2660 & 3.6368 & 10.7628 & 3.7318 & \textbf{2.5035}\\
       & VKDRS & 7.9710 & 5.2025 & 3.5941 & 10.6935 & 3.7277 & \textbf{3.0748}\\
      \midrule
      \multirow{6}{*}{CI} & ATFTIT & -173.5327 & -135.2131 & -245.1589 & 230.8767 & -220.9187 & \textbf{-649.3339}\\
       & DGPHGM & -211.9219 & -161.1084 & -78.7968 & 256.3937 & -427.6536 & \textbf{-650.7728}\\
       & GIKAVM & -229.8016 & -142.1957 & -141.2527 & 237.0655 & -428.5189 & \textbf{-642.3317}\\
       & IQFHFH & -168.0082 & -218.2147 & -190.1700 & 260.4627 & -332.3845 & \textbf{-644.8488}\\
       & LRKANG & -238.0650 & -203.9761 & -179.3537 & 245.8781 & -328.4392 & \textbf{-642.6191}\\
       & SIHSYS & -239.1328 & -248.0537 & -138.7962 & 253.7515 & -345.3256 & \textbf{-427.7065}\\
      \bottomrule
    \end{tabular}}
  \end{minipage}
  \hfill
  \begin{minipage}[t]{0.48\textwidth}
    \centering
    \textbf{(d) Pairwise Entanglement}
    \vspace{2mm}

    \resizebox{\linewidth}{!}{
    \begin{tabular}{ll| cccc|cc}
      \toprule
        \multicolumn{2}{c|}{{Protein}} & \multicolumn{4}{c|}{{Baselines}} & \multicolumn{2}{c}{\textbf{Ours}} \\
   \cmidrule(lr){1-2} \cmidrule(lr){3-6} \cmidrule(lr){7-8}
      Motifs & \makecell{Protein\\Sequences} & \makecell{SeqVQE} & \makecell{HBatch} & \makecell{EQC} & \makecell{CVaR-\\VQE} & \makecell{Graph-VQE\\no\_Global} & \makecell{\textbf{Graph-}\\\textbf{VQE}} \\
      \midrule
      \multirow{4}{*}{SI} & KPFKF & 5.1530 & 7.4570 & \textbf{3.7434} & 10.2478 & 4.2897 & 4.2848\\
       & NIGGF & 6.6943 & 6.4076 & \textbf{3.7448} & 9.7836 & 4.2654 & 4.2360\\
       & RYRDV & 6.6429 & 4.1322 & \textbf{3.7465} & 9.3523 & 4.2574 & 4.3208\\
       & VKDRS & 5.5267 & 4.6344 & \textbf{3.7440} & 9.8874 & 4.3019 & 4.2834\\
      \midrule
      \multirow{6}{*}{CI} & ATFTIT & -594.9252 & -586.3999 & -382.2423 & -145.5736 & -335.8122 & \textbf{-647.2968}\\
       & DGPHGM & -557.4127 & -606.7237 & -329.9556 & -94.4938 & -329.6120 & \textbf{-645.3686}\\
       & GIKAVM & -567.9143 & -489.1028 & -339.2926 & -131.0537 & -518.4131 & \textbf{-643.1144}\\
       & IQFHFH & -500.3024 & -460.5344 & -430.9311 & -5.5502 & -363.6979 & \textbf{-650.7102}\\
       & LRKANG & -528.9732 & -583.7809 & -396.6373 & -224.9334 & -627.1621 & \textbf{-648.9207}\\
       & SIHSYS & -528.8795 & -535.3878 & -269.2233 & -72.4734 & -270.2695 & \textbf{-644.0627}\\
      \bottomrule
    \end{tabular}}
  \end{minipage}

  \vspace{4mm}

  \begin{minipage}[t]{0.48\textwidth}
    \centering
    \textbf{(e) Reverse Linear Entanglement}
    \vspace{2mm}

    \resizebox{\linewidth}{!}{
    \begin{tabular}{ll| cccc|cc}
      \toprule
        \multicolumn{2}{c|}{{Protein}} & \multicolumn{4}{c|}{{Baselines}} & \multicolumn{2}{c}{\textbf{Ours}} \\
   \cmidrule(lr){1-2} \cmidrule(lr){3-6} \cmidrule(lr){7-8}
      Motifs & \makecell{Protein\\Sequences} & \makecell{SeqVQE} & \makecell{HBatch} & \makecell{EQC} & \makecell{CVaR-\\VQE} & \makecell{Graph-VQE\\no\_Global} & \makecell{\textbf{Graph-}\\\textbf{VQE}} \\
      \midrule
      \multirow{4}{*}{SI} & KPFKF & 7.0097 & 8.1459 & 3.7463 & 10.4262 & \textbf{2.7852} & 4.2488\\
       & NIGGF & 6.5474 & 8.1310 & 3.7473 & 10.2068 & \textbf{2.8334} & 3.7646\\
       & RYRDV & 8.2747 & 7.7544 & 3.7480 & 10.8743 & 2.6161 & \textbf{2.5070}\\
       & VKDRS & 3.0734 & 5.3438 & 3.7468 & 10.3171 & 2.5076 & \textbf{2.4985}\\
      \midrule
      \multirow{6}{*}{CI} & ATFTIT & -295.3364 & -203.9902 & -238.8986 & 193.3379 & -644.4987 & \textbf{-647.9050}\\
       & DGPHGM & -248.7907 & -241.2942 & -198.9152 & 218.2727 & -581.6541 & \textbf{-647.8729}\\
       & GIKAVM & -276.3248 & -249.7829 & -215.2232 & 237.8194 & \textbf{-649.9535} & -408.8670\\
       & IQFHFH & -237.4680 & -304.3901 & -253.1349 & 260.4716 & -251.1944 & \textbf{-646.6115}\\
       & LRKANG & -303.2892 & -204.4619 & -342.4690 & 226.3477 & \textbf{-644.7650} & -575.0506\\
       & SIHSYS & -276.3195 & -307.9258 & -233.7955 & 264.0772 & -56.7222 & \textbf{-653.6226}\\
      \bottomrule
    \end{tabular}}
  \end{minipage}
  \hfill
  \begin{minipage}[t]{0.48\textwidth}
    \centering
    \textbf{(f) SCA Entanglement}
    \vspace{2mm}

    \resizebox{\linewidth}{!}{
    \begin{tabular}{ll| cccc|cc}
      \toprule
        \multicolumn{2}{c|}{{Protein}} & \multicolumn{4}{c|}{{Baselines}} & \multicolumn{2}{c}{\textbf{Ours}} \\
   \cmidrule(lr){1-2} \cmidrule(lr){3-6} \cmidrule(lr){7-8}
      Motifs & \makecell{Protein\\Sequences} & \makecell{SeqVQE} & \makecell{HBatch} & \makecell{EQC} & \makecell{CVaR-\\VQE} & \makecell{Graph-VQE\\no\_Global} & \makecell{\textbf{Graph-}\\\textbf{VQE}} \\
      \midrule
      \multirow{4}{*}{SI} & KPFKF & 9.1162 & 8.7335 & 3.7474 & 10.4863 & 3.1369 & \textbf{3.1079}\\
       & NIGGF & 9.3350 & 8.0611 & 3.7483 & 10.5554 & 3.3978 & \textbf{3.0869}\\
       & RYRDV & 8.4991 & 6.7120 & 3.7459 & 10.6514 & 3.4216 & \textbf{3.1015}\\
       & VKDRS & 9.4542 & 9.9060 & 3.7477 & 10.6154 & \textbf{3.1100} & 4.8320\\
      \midrule
      \multirow{6}{*}{CI} & ATFTIT & -175.5876 & -235.1307 & -257.6591 & 267.3605 & -642.6381 & \textbf{-646.8710}\\
       & DGPHGM & -182.6266 & -122.8900 & -233.1663 & 214.1675 & -320.9747 & \textbf{-650.0346}\\
       & GIKAVM & -171.6665 & -163.3034 & -327.1881 & 215.2099 & \textbf{-426.6658} & -325.5992\\
       & IQFHFH & -191.4354 & -238.4471 & -268.0437 & 195.6143 & -327.7270 & \textbf{-427.9232}\\
       & LRKANG & -284.5228 & -209.3618 & -251.3855 & 80.4588 & \textbf{-646.7756} & -336.5671\\
       & SIHSYS & -179.8633 & -190.0826 & -280.8959 & 225.2807 & -622.5292 & \textbf{-647.7456}\\
      \bottomrule
    \end{tabular}}
  \end{minipage}

  \vspace{2mm}
  \parbox{\textwidth}{\footnotesize\textbf{Note:} SI = Stabilizing Interaction; CI = Competing Interaction. Bold = lowest energy among all methods.}
\end{table*}

\subsection{Convergence Analysis}
\begin{figure}[htbp]
  \centering
  \resizebox{\linewidth}{!}{\includegraphics{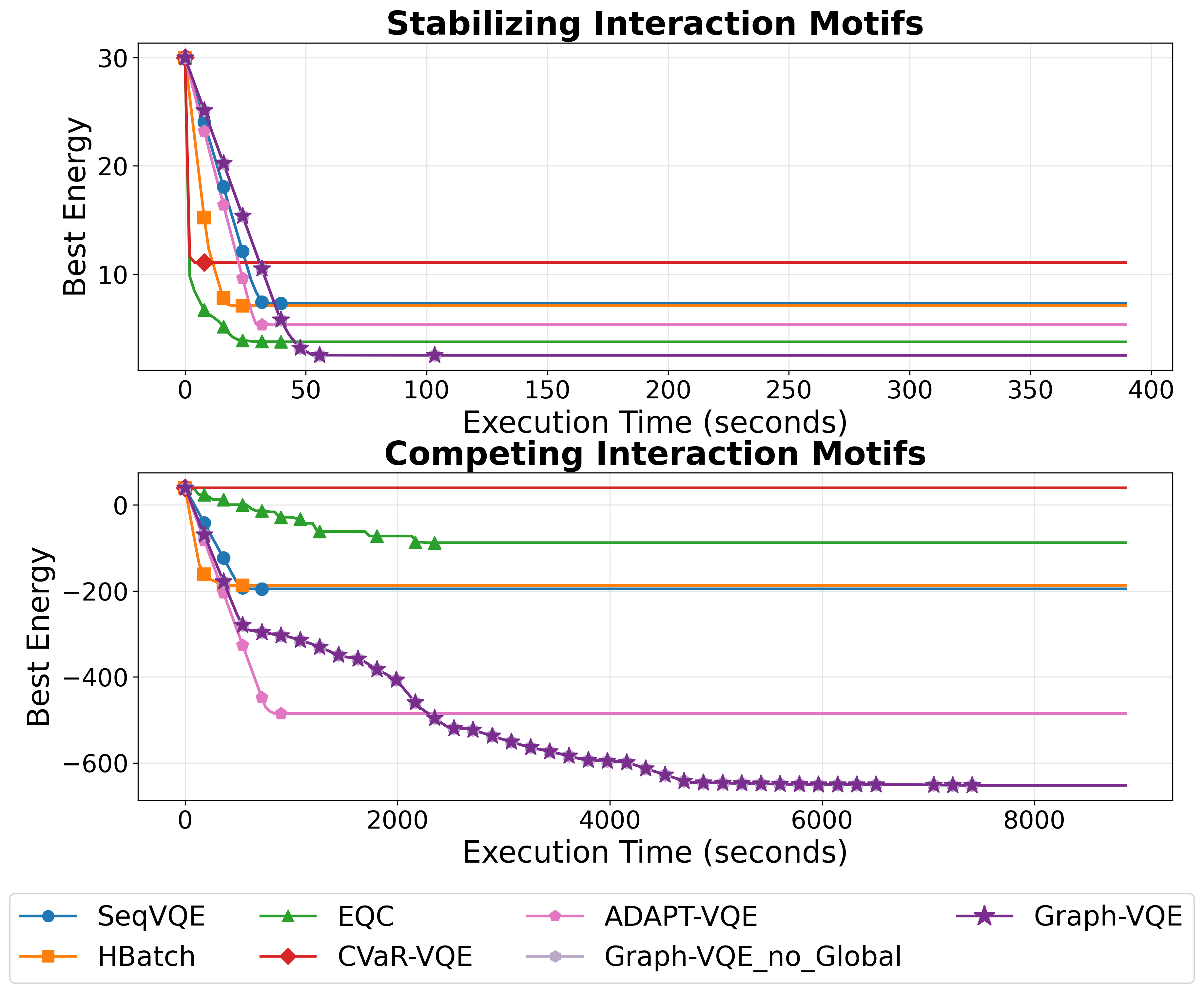}}
  \caption{Lowest Energy vs Execution Time, averaged over all evaluated sequences using 4 GPUs (Circular Entanglement). Lower energy ($\downarrow$) indicates better optimization. Left: SI Motifs. Right: CI Motifs.}
  \label{fig:energy_vs_time}
\end{figure}

To test whether baselines can eventually match Graph-VQE given more time, all methods are run with a time budget equal to Graph-VQE's convergence time. Figure~\ref{fig:energy_vs_time} tracks the best energy over execution time, averaged across all instances per motif category; markers indicate improvement points where a new best energy is discovered. For stabilizing interaction motifs (left), Graph-VQE descends steeply to a final energy around $2.5$, while EQC and ADAPT-VQE plateau at $3.7$--$5.0$ and SeqVQE/HBatch stall at higher values. The gap widens for competing interaction motifs (right): Graph-VQE reaches below $-640$, whereas SeqVQE and HBatch settle around $-200$ to $-300$ despite the full time budget, and CVaR-VQE remains trapped at positive energies indicating invalid structures. In both regimes, baseline curves flatten well before Graph-VQE's energy level, confirming that Hamiltonian-aware decomposition changes the optimization trajectory rather than merely reducing wall-clock cost.

For every peptide, C$\alpha$ coordinates from the experimental receptor segment in the benchmark complex (PDBbind pocket and ligand definitions) define the reference frame; RMSD is the optimal rigid superposition residual (Kabsch/SVD alignment via Biopython) between that reference and the predicted backbone, in \AA\@.
IBM quantum processor baselines use the VQE-predicted structure from the hardware baseline (full PDB C$\alpha$ trace), while AlphaFold3 baselines use the top-ranked mmCIF model, so both baselines are compared to the same crystallographic segment.
Binding affinity is the mean AutoDock Vina score (kcal/mol) over repeated docking trials: the receptor is the predicted full-atom model (AlphaFold3: CIF$\rightarrow$PDB; Graph-VQE: C$\alpha$ trace from our XYZ relaxed with Modeller, then PDBQT), the ligand is the cognate from the same complex with center-of-mass centered at the origin, and a $18\times18\times18$~\AA\ search box is centered on the ligand; lower (more negative) affinity indicates a stronger predicted pose.
Tables~\ref{tab:affinity_comparison} and~\ref{tab:rmsd_comparison} summarize the best value each method attains per sequence; the Graph-VQE column picks the entanglement layout (Best Ent.) that minimizes RMSD or affinity on that row, while baselines are single reported structures. Across both tables, Graph-VQE is most often best on alignment and binding among the three columns, and competitive with the hardware baseline where both exist, which supports that Hamiltonian-aware optimization is recovering physically plausible pocket geometries rather than only improving internal energy.

\begin{table}[htbp]
  \caption{Best Binding Affinity Comparison (kcal/mol, lower is better)}
  \label{tab:affinity_comparison}
  \centering
  \footnotesize

  \resizebox{\linewidth}{!}{
  \begin{tabular}{ll| cc|cc}
    \toprule
      \multicolumn{2}{c|}{{Protein}} & \multicolumn{2}{c|}{{Baselines}} & \multicolumn{2}{c}{\textbf{Ours}} \\
    \cmidrule(lr){1-2} \cmidrule(lr){3-4} \cmidrule(lr){5-6}
    Motifs & \makecell{Protein\\Sequences} & \makecell{IBM\\quantum processors} & \makecell{AlphaFold3} & \makecell{\textbf{Graph-}\\\textbf{VQE}} & \makecell{Best\\Ent.} \\
    \midrule
    \multirow{4}{*}{SI} & KPFKF & -- & $-3.878$ & $\mathbf{-3.976}$ & circular \\
     & NIGGF & $-3.459$ & $-3.613$ & $\mathbf{-3.864}$ & linear \\
     & RYRDV & $-4.868$ & $-4.503$ & $\mathbf{-5.424}$ & pairwise \\
     & VKDRS & $\mathbf{-3.524}$ & $-3.329$ & $-3.511$ & linear \\
    \midrule
    \multirow{6}{*}{CI} & ATFTIT & $-2.307$ & $-2.113$ & $\mathbf{-2.769}$ & sca \\
     & DGPHGM & $-3.138$ & $\mathbf{-3.649}$ & $-3.436$ & sca \\
     & GIKAVM & $-3.532$ & $-3.396$ & $\mathbf{-3.782}$ & pairwise \\
     & IQFHFH & $-3.013$ & $-3.073$ & $\mathbf{-3.172}$ & sca \\
     & LRKANG & $-3.832$ & $-3.666$ & $\mathbf{-4.450}$ & full \\
     & SIHSYS & $-3.581$ & $-3.382$ & $\mathbf{-4.273}$ & linear \\
    \bottomrule
  \end{tabular}}

  \vspace{2mm}
  \parbox{\linewidth}{\footnotesize\textbf{Note:} SI = Stabilizing Interaction; CI = Competing Interaction. Bold = best (most negative) affinity among all methods.}
\end{table}

\begin{table}[htbp]
  \caption{Best RMSD Comparison (\AA, lower is better)}
  \label{tab:rmsd_comparison}
  \centering
  \footnotesize

  \resizebox{\linewidth}{!}{
  \begin{tabular}{ll| cc|cc}
    \toprule
      \multicolumn{2}{c|}{{Protein}} & \multicolumn{2}{c|}{{Baselines}} & \multicolumn{2}{c}{\textbf{Ours}} \\
    \cmidrule(lr){1-2} \cmidrule(lr){3-4} \cmidrule(lr){5-6}
    Motifs & \makecell{Protein\\Sequences} & \makecell{IBM\\quantum processors} & \makecell{AlphaFold3} & \makecell{\textbf{Graph-}\\\textbf{VQE}} & \makecell{Best\\Ent.} \\
    \midrule
    \multirow{4}{*}{SI} & KPFKF & -- & $\mathbf{0.553}$ & $1.154$ & full \\
     & NIGGF & $2.000$ & $1.963$ & $\mathbf{1.699}$ & circular \\
     & RYRDV & $2.061$ & $1.683$ & $\mathbf{1.198}$ & sca \\
     & VKDRS & $0.831$ & $1.688$ & $\mathbf{0.513}$ & circular \\
    \midrule
    \multirow{6}{*}{CI} & ATFTIT & $1.655$ & $\mathbf{1.317}$ & $1.347$ & sca \\
     & DGPHGM & $2.322$ & $2.541$ & $\mathbf{1.786}$ & circular \\
     & GIKAVM & $2.890$ & $3.664$ & $\mathbf{1.893}$ & pairwise \\
     & IQFHFH & $1.914$ & $\mathbf{0.641}$ & $1.653$ & full \\
     & LRKANG & $2.911$ & $\mathbf{0.978}$ & $1.660$ & pairwise \\
     & SIHSYS & $2.846$ & $1.758$ & $\mathbf{0.949}$ & circular \\
    \bottomrule
  \end{tabular}}

  \vspace{2mm}
  \parbox{\linewidth}{\footnotesize\textbf{Note:} SI = Stabilizing Interaction; CI = Competing Interaction. Bold = best (lowest) RMSD among all methods.}
  \vspace{-0.1in}
\end{table}

\section{Discussion}
\label{sec:discussion}

Having demonstrated Graph-VQE's accuracy and scaling advantages, we now discuss the simulation backend trade-offs and practical considerations that shaped the framework's design.

\subsection{Simulation Backend Trade-offs}
\begin{figure}[htbp]
  \centering
  \resizebox{\linewidth}{!}{\includegraphics{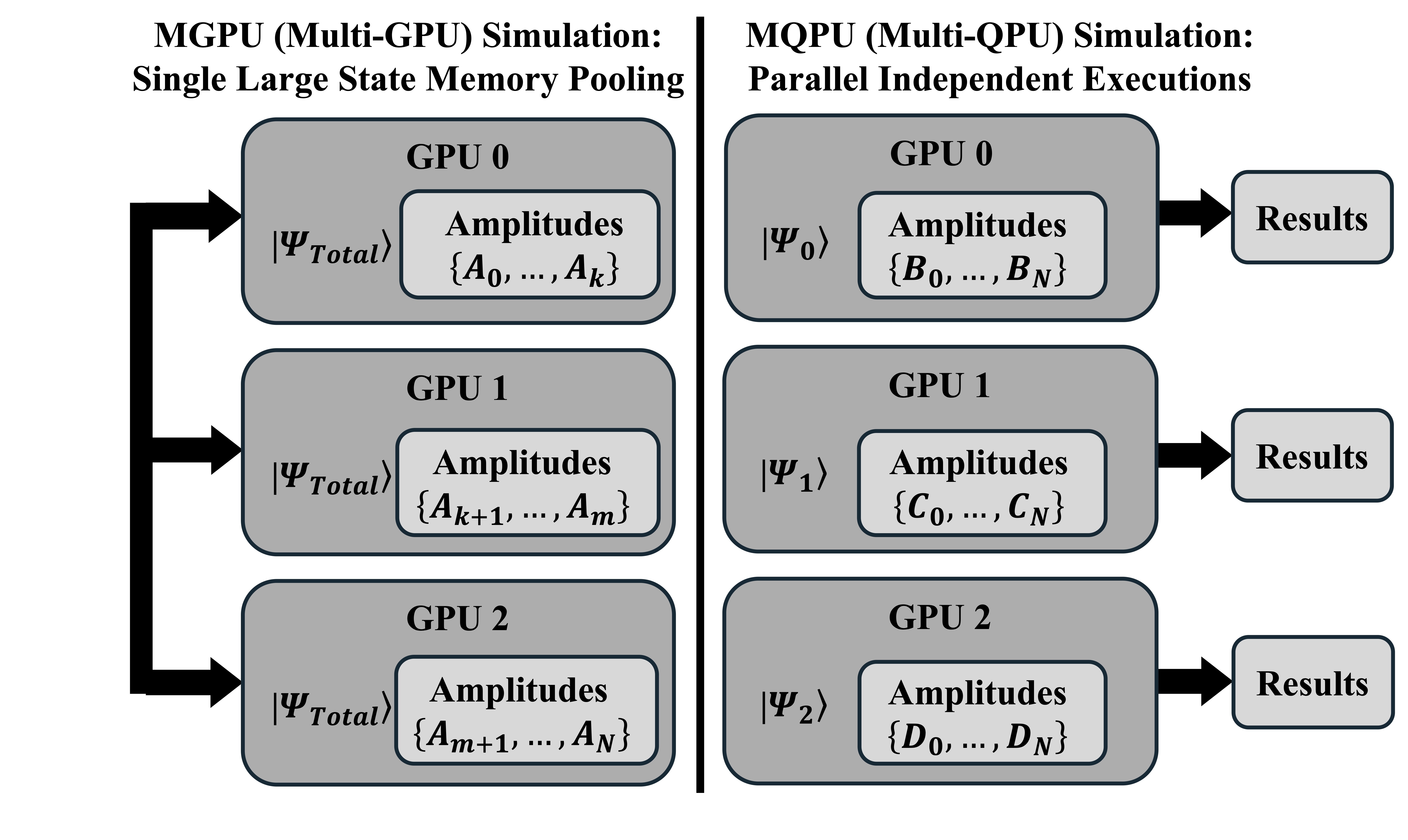}}
  \caption{Comparison of CUDA-Q backends: multi-GPU memory pooling (left) vs. multi-QPU simulation (right).}
  \label{fig:mgpu_vs_mqpu}
\end{figure}

This work uses the CUDA-Q multi-QPU simulation backend, which treats each GPU as an independent virtual QPU for parallel circuit execution~\cite{the_cuda_q_development_team_2026_19057431}, directly supporting both Graph-VQE's task parallelism (block optimization) and data parallelism (Hamiltonian batching). An alternative backend, multi-GPU memory pooling, pools memory across GPUs via MPI to simulate a single larger state vector, addressing memory constraints for circuits exceeding single-GPU capacity. However, as illustrated in Figure~\ref{fig:mgpu_vs_mqpu}, these two backends cannot be enabled simultaneously: multi-GPU memory pooling assumes all GPUs cooperatively store fragments of one shared state $|\Psi_{\text{Total}}\rangle$ with coordinated inter-GPU updates, whereas multi-QPU simulation treats each GPU's state as fully independent. Enabling both would create irreconcilable assumptions about amplitude ownership, producing incorrect results.

CUDA-Q also offers tensor network backends (e.g., Matrix Product State), which can handle large qubit counts by exploiting low-entanglement structure~\cite{schieffer2025harnessing}. However, bond dimension truncation discards quantum correlations, and VQE circuits with EfficientSU2 at depth~4 generate substantial entanglement that MPS struggles to represent compactly~\cite{hubig2018error, alvertis2025classical}. Graph-VQE instead uses exact state vector simulation so that the only source of variance is measurement shot noise rather than simulation approximation artifacts.

\subsection{Optimization Under Measurement Noise}
Graph-VQE is designed for finite-shot evaluation where each energy estimate is subject to statistical noise scaling as $\mathcal{O}(1/\sqrt{N_{\text{shots}}})$. Block decomposition provides inherent noise resilience by reducing the dimensionality of each subproblem, limiting the opportunity for shot noise to mislead the optimizer. The subsequent global refinement benefits from the high-quality initialization produced by block optimization, requiring fewer iterations and thus accumulating less noise-induced error.

\subsection{Hyperparameter Configuration}
Graph-VQE exposes several hyperparameters whose trade-offs practitioners should be aware of. The Louvain resolution $\gamma$ controls block granularity: higher values yield smaller, faster-to-optimize blocks at the risk of missing inter-block correlations; lower values capture more interactions but increase per-block difficulty. Ansatz depth (reps$=$4 by default) trades expressibility against barren plateaus and runtime. Measurement shots (8192) balance estimator variance against cost. Early stopping (tolerance $10^{-4}$, patience 12, window 40) and epoch patience (3) prevent both premature termination and wasted computation. As a practical guideline: increase epoch patience if blocks converge to poor energies, reduce ansatz depth or add GPUs if optimization is slow, and increase shots if energy estimates appear noisy.

\section{Conclusion}
\label{sec:conclusion}

This paper presents Graph-VQE, a biologically-informed framework that scales VQE by partitioning molecular Hamiltonians into weakly-coupled blocks via Louvain community detection. By combining block-wise restricted updates on the full-Hamiltonian objective with a short global refinement stage, Graph-VQE reaches lower energy states than standard global optimization methods while scaling to additional GPUs with decreasing wall-clock time and preserved solution quality. On a protein folding benchmark, the framework navigates into deep energy regions unreachable by baselines under the same settings, achieving competitive RMSD and binding affinity relative to AlphaFold3 and IBM quantum processor baselines, offering a practical pathway for high-fidelity quantum simulations in drug discovery and protein pathology research. The Qiskit-CUDA-Q adaptation layer further makes these gains accessible within familiar quantum application workflows. More broadly, the results suggest that Hamiltonian-aware optimization-level parallelism is a promising route for scaling hybrid quantum-classical biomolecular simulation on emerging multi-QPU platforms. The benchmark results further indicate that this strategy can improve both optimization quality and time-to-solution without relying on costly circuit cutting or exponential classical recombination. Future work will extend the framework beyond coarse-grained lattice models and investigate larger biomolecular systems with higher qubit demands.

\newpage
\bibliographystyle{IEEEtran}
\bibliography{refs}

\end{document}